\documentclass[a4paper,fleqn,usenatbib]{mnras}
\usepackage{newtxtext,newtxmath}
\usepackage[T1]{fontenc}
\usepackage{ae,aecompl}

\usepackage{graphicx}   
\usepackage{amsmath}    
\usepackage{amssymb}    

\newcommand{\cii}{\ion{C}{ii}}
\newcommand{\siiv}{\ion{Si}{iv}}
\newcommand{\mgii}{\ion{Mg}{ii}}
\newcommand{\civ}{\ion{C}{iv}}
\newcommand{\oiv}{\ion{O}{iv}}

\title[Spectroscopic and imaging observations of small-scale reconnection events]{Spectroscopic and imaging observations of small-scale reconnection events}

\author[Dong~Li ]{
Dong~Li,$^{1,2,3}$\thanks{E-mail:lidong@pmo.ac.cn}, Leping~Li$^{2}$, and Zongjun~Ning$^{1}$ \\
$^{1}$Key Laboratory for Dark Matter and Space Science, Purple
Mountain Observatory, CAS, Nanjing 210034, China \\
$^{2}$CAS Key Laboratory of Solar Activity, National Astronomical
Observatories, Beijing 100012, China \\
$^{3}$State Key Laboratory of Space Weather, Chinese Academy of
Sciences, Beijing 100190, China \\}

\date{Accepted 2018 June 26. Received 2018 June 26; in original form 2018 March 23}

\pubyear{2018}
\begin{document}
\label{firstpage}
\pagerange{\pageref{firstpage}--\pageref{lastpage}}
\maketitle

\begin{abstract}
We present spectroscopic and imaging observations of small-scale
reconnection events on the Sun. Using the Interface Region Imaging
Spectrograph (IRIS) observations, one reconnection event is first
detected as IRIS jets with fast bi-directional velocities in the
chromosphere and transition region, which are identified as
non-Gaussian broadenings with two extended wings in the line
profiles of \siiv, \cii, and \mgii~k. The magnetograms under the
IRIS jets from Helioseismic and Magnetic Images exhibit magnetic
flux cancellation simultaneously, supporting that the IRIS jets are
driven by magnetic reconnection. The Atmospheric Imaging Assembly
images also show an extreme ultraviolet (EUV) brightening which is
shortly after the underlying IRIS jets, i.e., in the 131~{\AA},
171~{\AA}, 193~{\AA}, 211~{\AA}, and 94~{\AA} channels, implying
that the overlying EUV brightening in the corona is caused by the
IRIS jets in the chromosphere and transition region. We also find
another three reconnection events which show the same features
during this IRIS observation. Our observational results suggest that
the small-scale reconnection events might contribute to the coronal
heating. The new result is that the process of magnetic reconnection
is detected from the photosphere through chromosphere and transition
region to the corona.
\end{abstract}


\section{Introduction}
Magnetic reconnection is a fundamental dynamical process in all
magnetized astrophysical plasmas including solar atmosphere. It is
thought to be the most efficient transfer of magnetic energy into
plasma bulk acceleration or heating particles to the higher energy
\citep{Priest02,Mackay12}. Therefore, it plays a significant role on
the solar eruptions in a wide range of scales from $\sim$1000~Mm to
$\sim$1~Mm, i.e., coronal mass ejections \citep[CMEs,][]{Lin05},
filament eruptions \citep{Mackay10}, solar flares
\citep{Aschwanden02}, bright points \citep[BPs,][]{Priest94}, and
transition-region explosive events \citep{Innes97}. In
two-dimensional (2D) reconnection model \citep{Shibata99,Priest00},
it occurs at an X-point where anti-parallel magnetic field lines
converge and reconnect. In the last few decades, using the high
temporal and spatial resolution data, many observational evidences
of magnetic reconnection on the Sun have been reported, i.e.,
loop-top hard X-ray source \citep{Masuda94}, cusp-shaped post-flare
loops\citep{Tsuneta92}, reconnection inflows or outflows
\citep{Yokoyama01,Takasao12}, magnetic null point \citep{Zhang12},
bi-directional moving structures \citep{Ning14}, and high-energetic
particles \citep{Klassen05,Li07}, as well as the bi-directional jets
with high velocity \citep{Dere91,Innes15}.

Using the extreme ultraviolet (EUV) and X-ray observations,
\cite{Su13} presented a process of magnetic reconnection during a
solar flare, which was observed when a set of coronal loops were
moving toward another set of loops and then formed an X-type
structure. The X-shape geometry can also be found in two
anti-parallel loops with small scale \citep{Yang15} or between an
eruption filament and its nearby coronal loops \citep{Lil16}. Many
signatures of magnetic reconnection were observed in a spectacular
eruptive flare (X8.2), such as the current sheets, reconnection
inflow, hot cusp structure, and post-flare loops \citep{Li18,Yan18}.
These observations are based on the high-resolution imaging data
\citep[e.g.,][]{Masuda94,Liu10,Xue16}, such as the Reuven Ramaty
High Energy Solar Spectroscopic Imager (RHESSI), the Atmospheric
Imaging Assembly (AIA) and Helioseismic and Magnetic Imager (HMI) on
board Solar Dynamics Observatory (SDO), and the New Vacuum Solar
Telescope (NVST). On the other hand, the bi-directional plasma jets
observed from the spectroscopic data usually considered as the
signature of small-scale magnetic reconnection
\citep{Dere94,Innes97}. They often appeared as non-Gaussian
broadenings with two extended wings of spectral lines, the extended
velocities could be more than 100~km~s$^{-1}$ in both red and blue
wings \citep{Brueckner83,Chae98,Curdt12}. Using the spectroscopic
observations from Solar Ultraviolet Measurements of Emitted
Radiation (SUMER) and Interface Region Imaging Spectrograph (IRIS),
a typical line profiles (i.e., \siiv, \oiv) with non-Gaussian
broadenings were found, which were produced by small-scale magnetic
reconnections \citep{Winebarger02,Innes15}. These small-scale
reconnections often appeared repetitively on the Sun
\citep{Doyle06,Gupta15}. Besides these imaging and spectroscopic
observations, the numerical results from magnetohydrodynamic (MHD)
simulations \citep[e.g.,][]{Roussev01,Heggland09,Ni15} also support
the fact that magnetic reconnection plays an important role in solar
eruptions.

On solar surface, the temperature in the corona ($\sim$10$^6$~K) is
much higher than that in the photosphere ($\sim$6$\times$10$^3$~K)
and chromosphere ($\sim$10$^4$~K). Moreover, the solar corona has
radiative losses by EUV emission, which indicates that the plasma
temperature could be maintained continuously by some heating source.
\citep{Aschwanden05}. However, it is still unclear how radiation
from such hot plasmas in the corona is maintained, which is famous
as `coronal heating problem' \citep{Klimchuk06,Reale14}. Until now,
various models have been proposed to account for this topic, such as
the dissipation of Alfv\'{e}n waves
\citep[e.g.,][]{McIntosh11,Mathioudakis13,Grant18}, the magnetic
energy released from the nanoflares
\citep{Parnell00,Antolin08,Ishikawa17} or the coronal BPs
\citep[CBPs,][]{Longcope98,Zhang01,Schmelz13}, and also the
small-scale but fast-moving jets \citep{Dep11,Tian14,Henriques16}.
For those models, the key step is to identify the origin of energy
release \citep{Klimchuk06}, which might be originally driven by the
constant photospheric motions, such as the movements of
intergranular lanes \citep{Ji12,Hong17}, or the flux cancellation of
magnetic fields \citep{Priest94,Innes13,Tian18}.

Using high-resolution observations in soft X-ray (SXR) and EUV
channels, various small-scale eruptions on the Sun have been
explored to discover the signatures of coronal heating process, such
as CBPs, nanoflares, transition-region explosive events, blinkers,
fast-moving jets
\citep[e.g.,][]{Benz98,Parnell02,Schmelz13,Yanl18,Young18}. CBPs are
small-scale brightness enhancements observed in SXR and EUV imaging
observations, their lifetime could be ranging from minutes to hours
\citep{Hong14,Alipour15}. Nanoflares are tiny brightenings in EUV
wavelengths, and their typical time scales are a few minutes
\citep{Benz98,Ishikawa17}. Transition-region explosive events are
small-scale and short-time transients observed in EUV lines in the
temperature ranges of about 6$\times$10$^{4}$~K $-$
7$\times$10$^{5}$~K, and these line profiles exhibit non-Gaussian
broadenings with two extended wings
\citep{Brueckner83,Dere91,Huang14}. Blinkers are small-scale
intensity enhancements in EUV lines
\citep{Harrison97,Chae00,Brkovic04}. The fast-moving jets are also
small scale with speeds of hundreds of km~s$^{-1}$, and their
lifetimes are around 60 s \citep{Tian14,Henriques16}. Until now, the
physical nature of these small-scale events is not yet fully
understood. However, they are believed to be related with the
movements of magnetic fields in the photosphere, i.e., flux
cancelling, emerging flux \citep[see][]{Dere91,Chae98,Innes13}.

It is well known that the origin of coronal heating lies in
photospheric energy, but how this gets to the corona is still an
open question. In this paper, using the spectroscopic and imaging
observations from the IRIS \citep{Dep14} and SDO \citep{Pesnell12},
we present a detailed investigation of four small-scale reconnection
events in solar atmosphere. Our observations support the magnetic
reconnection model, and might be helpful to coronal heating problem.

\section{Observations}
\begin{figure}
\includegraphics[width=\columnwidth]{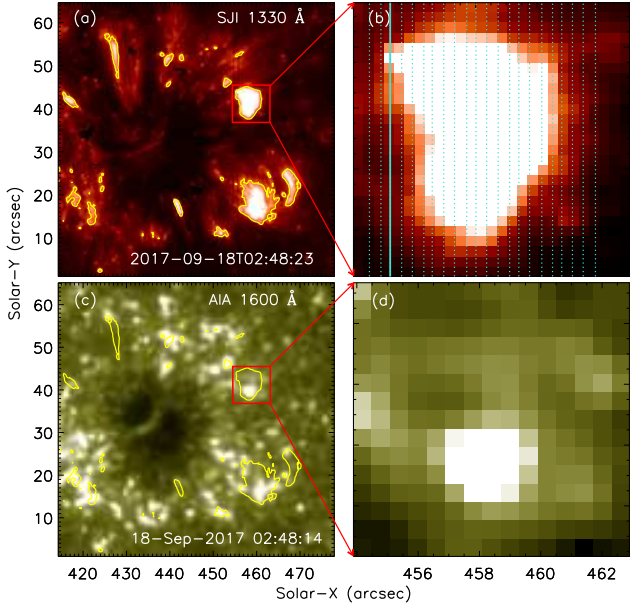}
\caption{Left: UV images in SJI~1330~{\AA} (a) and AIA~1600~{\AA}
(c). The yellow contours are from the SJI~1330~{\AA} intensities.
Right: Zoomed images ($\sim$8.7\arcsec$\times$8.7\arcsec) outlined
by the red box in SJI~1330~{\AA} (b) and AIA~1600~{\AA} (d). The
solid vertical line marks the slit of IRIS at the given time
($\sim$02:48:23~UT), while the dashed vertical lines mark the IRIS
slits at other times.} \label{image}
\end{figure}

IRIS performed two very large dense rasters in the NOAA active
region (AR) 12680 from 01:13:14~UT to 02:51:26~UT on 2017 September
18, i.e., 175\arcsec\ along the spectral slit, 320 raster steps with
a step size of $\sim$0.35\arcsec. Thus, the field-of-view (FOV) of
Slit-Jaw Imager (SJI) for this observation is about
112\arcsec$\times$175\arcsec with a time cadence of $\sim$37~s. Each
raster lasted for about 2946~s with a step cadence of $\sim$9.2~s.
Therefore, each SJI image contains four raster steps.
Figure~\ref{image} shows the context images from IRIS/SJI
\citep{Dep14} and SDO/AIA \citep{Lemen12} observations. Notice that
the AIA data have been ``re-spiked'' with the routine of
AIA\_RESPIKE.pro in solar software (SSW), as pervious observations
found that some EUV kernels were incorrectly flagged by the AIA
de-spiking routine \citep{Young13,Li16b}. Then the AIA~1600~{\AA}
image (0.6\arcsec/pixel) is used to co-align with the SJI~1330~{\AA}
image (0.33\arcsec/pixel) by cross-correlation since they both
include the UV continuum emissions, as indicated by the yellow
contours in panels~(a) and (c). The red box outlines a bright
feature in Far Ultraviolet (FUV) wavelength, it is located at the
edge of a sunspot. Panels~(b) and (d) give the zoomed images in
SJI~1330~{\AA} and AIA~1600~{\AA}. The FUV bright event is scanned
by the slits of IRIS, as shown by the vertical lines in panel~(b).
The solid vertical line marks the site of IRIS slit at around
02:48:23~UT, which is also the time of the given SJI~1330~{\AA}
image in panel~(a). Only the bottom part of the FUV bright event is
seen in the AIA~1600~{\AA} image, which might be due to the
different dominant emission lines in AIA~1600~{\AA} (\civ) and
SJI~1330~{\AA} (\cii), resulting the formed temperature in
AIA~1600~{\AA} \citep[log$T~\sim$~5.0,][]{Lemen12} is higher than
that in SJI~1330~{\AA} \citep[log$T~\sim$~4.3,][]{Dep14}.

\section{Results}
\begin{figure}
\includegraphics[width=\columnwidth]{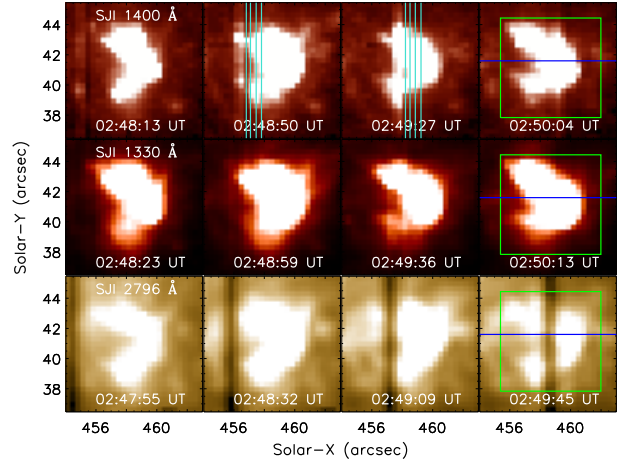}
\caption{Snapshots in SJI 1400~{\AA}, 1330~{\AA}, and 2796~{\AA}
with a FOV of about 8.7\arcsec$\times$8.7\arcsec. The vertical lines
mark the IRIS slits at the given snapshot. The green box outline the
regions that used for the calculation of the intensities and
magnetic fluxes shown in Figure~\ref{flux}. The horizontal lines
outline the loop-top region.} \label{snap}
\end{figure}

To investigate the evolution of this FUV bright event, a series of
FUV images in SJI~1400~{\AA} are shown in the upper panels of
Figure~\ref{snap}. The FUV bright event displays a loop-like
structure at these given panels. At the beginning
($\sim$02:48:13~UT), the bright event is not yet scanned by the
slits of IRIS. Then the bright event reaches its maximum brightness
at about 02:48:50~UT, and it is scanned by the IRIS slits. Finally,
the FUV bright event becomes weaker and weaker, i.e., at around
02:49:27~UT and 02:50:04~UT. During this time interval, the spectral
slits firstly go through the double loop legs, and then cross the
loop-top region (marked by the blue horizontal line), as indicated
by the vertical lines. As mentioned above, the time cadence of the
IRIS/SJI is only a quarter of IRIS spectra. Therefore, each SJI
image is over-plotted with four vertical lines which represent the
spectral slit sites. The middle panels show the similar FUV bright
event in SJI~1330~{\AA} images. The lower panels give a series of
Near Ultraviolet (NUV) images in SJI~2796~{\AA}, which also exhibit
the NUV bright event with a similar behaviors as that
in~SJI~1400~{\AA} and 1330~{\AA}.

\begin{figure}
\includegraphics[width=\columnwidth]{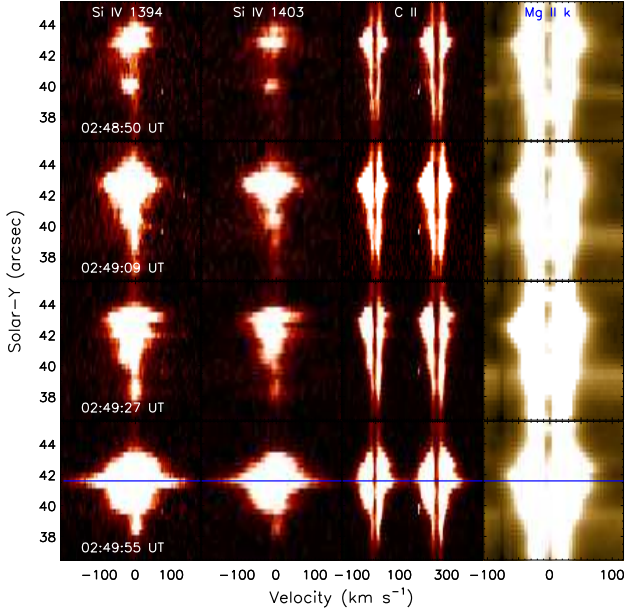}
\caption{Simultaneous IRIS spectra of the reconnection event at the
spectral windows of `\siiv~1394~{\AA}', `\siiv~1403~{\AA}', `\cii'
and `\mgii~k', respectively. The observed times are given in the
left panels. The horizontal lines outline the loop-top region in
Figure~\ref{snap}.} \label{spectra}
\end{figure}

Figure~\ref{spectra} shows the IRIS spectra during the FUV/NUV
bright event scanned by the IRIS slits at four IRIS windows, i.e.,
`\siiv~1394~{\AA}', `\siiv~1403~{\AA}', `\cii' and `\mgii~k'.
Although the bright event has achieved its maximum brightness in FUV
band at about 02:48:50~UT (Figure~\ref{snap}), the slit of IRIS
scans two loop legs of the FUV/NUV bright event, which exhibits two
clearly enhancements of the line cores in \siiv~1393.76~{\AA} and
1402.77~{\AA} (6.3$\times$10$^4$~K), \cii~1334.53~{\AA} and
1335.71~{\AA} (2$\times$10$^4$~K), i.e., the slit positions at
$\sim$39\arcsec\ and $\sim$43\arcsec. Then the slits of IRIS move
through the main body of the FUV/NUV bright event, the line cores
are continued to be enhanced. At the same time, two line wings are
also enhanced simultaneously, which can be extended to more than
100~km~s$^{-1}$ in both red and blue wings, as outlined by the
horizontal lines in the lowest panels in Figure~\ref{spectra}. These
extended red and blue wings are similar to the bi-directional jets
reported by \cite{Innes97}, which could be regarded as the
reconnection event. Thus the loop-top locations marked by the
horizontal lines are most likely the reconnection sites. We also
note that the blue/red wings of \mgii~k (10$^4$~K) line can also be
nearly 100~km~s$^{-1}$ (the lowest panel), indicating the
reconnection event might take place from the lower chromosphere to
the transition region. Therefore, it is different from the pervious
transition-region explosive events \citep[e.g.,][]{Innes97,Huang14},
which are usually detected in the spectral lines with formation
temperatures between 6$\times$10$^4$~K and 7$\times$10$^5$~K
\citep{Wilhelm07}.

\begin{figure}
\includegraphics[width=\columnwidth]{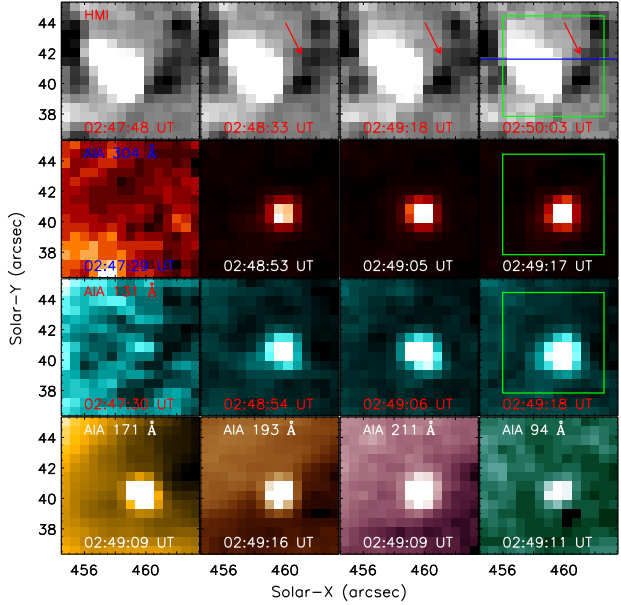}
\caption{HMI LOS magnetograms in scale levels of $\pm$50~G, as well
as AIA EUV images in 304~{\AA}, 131~{\AA}, 171~{\AA}, 193~{\AA}, and
94~{\AA}. The green box outline the regions that used for the
calculation of the intensities and magnetic fluxes shown in
Figure~\ref{flux}. The horizontal line outlines the loop-top region
in Figure~\ref{snap}.} \label{magnetic}
\end{figure}

To confirm this reconnection event, we then plot the line-of-sight
(LOS) magnetograms from SDO/HMI \citep{Schou12}, as shown in the
upper panels of Figure~\ref{magnetic}. We can see that the loop-top
region is located above the polarity inversion line between the
positive and negative magnetic fields, as indicated by the red arrow
and horizontal line. This observational fact suggests a possible
flux cancellation of opposite magnetic polarities
\citep{Innes13,Li16c}. On the other hand, the positive magnetic
fields during the reconnection event become stronger and stronger
(indicated by the red arrows), implying the flux emergence of
positive magnetic field during the reconnection event
\citep{Solanki03,Zhao17}. In contrary, the negative magnetic fields
become weaker and weaker, confirming the magnetic flux cancellation.
All these observational facts suggest that the evolution of magnetic
fields in the photosphere provide the energy for driving the IRIS
jets in the chromosphere and transition region
\citep{Peter14,Tian18}.

To examine the response of the reconnection event in the corona, we
also show the AIA images in EUV passbands in the rows 2$-$4 of
Figure~\ref{magnetic}. The first-column panels in the rows of 2 and
3 display the EUV images at around 02:47~UT in AIA~304~{\AA} and
131~{\AA}, respectively. They does not exhibit any bright features
at the reconnection event region outlined by the green box. However,
both of these two channels appear to brighten in the green box
region between $\sim$02:48~UT$-$02:49~UT. The other AIA EUV channels
also become bright simultaneously, i.e., in AIA~171~{\AA},
193~{\AA}, 211~{\AA}, and 94~{\AA}, as shown in the lowest panels in
Figure~\ref{magnetic}. The EUV brightening from the SDO/AIA is
situated closely to the reconnection sites, suggesting that the EUV
brightening in the corona is caused by the magnetic reconnection
which occurs in the chromosphere and transition region. All these
observational facts imply that the energy that used to heat the
solar corona originates from the magnetic flux cancellation in the
photosphere.

\begin{figure}
\includegraphics[width=\columnwidth]{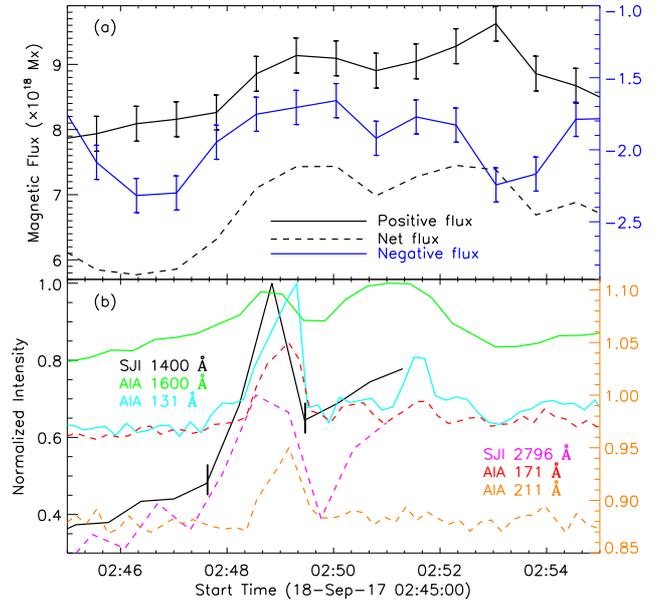}
\caption{Panel~(a): Temporal evolution of the positive (black),
negative (blue) and net (dashed) magnetic fluxes from the red box
given in Figure~\ref{magnetic}. The error bars ($\pm \sigma$) are
given for positive and negative fluxes, respectively. Panel~(b):
Time series of the normalized intensity in SJI~1400~{\AA} (black)
and 12796~{\AA} (purple), as well as AIA 1600~{\AA} (green),
131~{\AA} (turquoise), 171~{\AA} (red), and 211~{\AA} (orange). Two
vertical ticks enclose the lifetime of the IRIS jets in
SJI~1400~{\AA}.} \label{flux}
\end{figure}

The reconnection event is identified as the flux cancellation of
opposite magnetic fields in the photosphere, the IRIS jets with fast
bi-directional velocities in the chromosphere and transition region,
the EUV brightening in the corona. To look closely their temporal
relationships, the time series of these parameters are plotted in
Figure~\ref{flux}, the integral region is outlined with a green box
in Figures~\ref{snap} and \ref{magnetic}. Panel~(a) shows that the
positive and net magnetic fluxes increase rapidly from
$\sim$02:47~UT to $\sim$02:49~UT, and then decrease. While the
negative magnetic fluxes are contrary, they greatly decrease and
then grow slowly. The standard deviations ($\sigma$) of positive and
negative magnetic fluxes are obtained from their time-evolution
profiles, as shown by the error bars. The maximum values of the
changing positive and negative fluxes are greater than their
standard deviations, indicating that the changed fluxes of magnetic
fields are reliable. Our observational results suggest that the rate
of positive flux emergence is greater than that of flux
cancellation. At the same time, the normalized intensities in
SJI~1400~{\AA} (black) and SJI~2796~{\AA} (purple) exhibit a
pronounced peak at almost the same time, as shown in panel~(b). We
note that double peaks appear in AIA 1600~{\AA} flux (green),
corresponding with two peaks in positive magnetic flux. However, the
second peak is missed by the IRIS observations. Therefore, only the
first one which corresponds to the SJI peak is studied in this
paper. The same peak are also detected in EUV passbands, such as
AIA~131~{\AA} (turquoise), 171~{\AA} (red), and 211~{\AA} (orange).
Notice that the solid lines are shown in the left axis, while the
dashed lines are displayed with the right axis. It is interesting
that the peak time of these EUV fluxes in SDO/AIA images is later
than that of the FUV/NUV fluxes in IRIS/SJI images, indicating that
the released energy in the chromosphere and transition region can
spread to the overlying corona. This observational result is
consistent with the traditional one-dimensional model of the solar
atmosphere \citep{Vernazza76,Vernazza81}.

\section{Discussions}
To clearly illustrate that the magnetic energy in the photosphere is
strong enough to drive the IRIS jets in the chromosphere and
transition region and then cause the EUV brightening in the corona,
the energy budgets in the solar atmospheres are given. Firstly, we
estimate the released magnetic energy (E$_m~=~E_1-E_2$) that derived
from photospheric magnetic fields. Here, E$_1$ and E$_2$ are the
total magnetic energy at the onset ($\sim$02:47~UT) and peak
($\sim$02:49~UT) times of reconnection event, respectively. They are
estimated by integrating $B^2/8\pi$ over the area (green box) shown
in the upper panels in Figure~\ref{magnetic} with a height of
$\sim$1~Mm \citep{Sharykin17}, which is the low-limit site of this
reconnection event, i.e., the height of lower chromosphere
\citep{Sturrock86}. Then considering the uncertainties of magnetic
fluxes (error bars in Figure~\ref{flux}~a), the released magnetic
energy is estimated to be about (6.7$\pm$1.9)$\times$10$^{27}$~erg.
Next, we can estimate the kinetic energy (E$_k~=~\rho V v^2$) in the
transition regions from the IRIS jets. Here, $\rho$ is the
transition-region density ($\sim$10$^{-13}$~g~cm$^{-3}$) on the Sun,
$v$ is the Doppler velocity ($\sim$150~km~s$^{-1}$) of
\siiv~1393.76~{\AA} line which formed in the transition region.
$\frac{1}{2}$ is removed from the kinetic energy as the IRIS jets
are bi-directional. The volume ($V~\approx~w^2~vt$) might be
estimated by the width ($w~\approx$~1\arcsec) and propagation
distance ($vt$) along the LOS of the IRIS jets at its maximum speed
in Figures~\ref{spectra} (bottom panel) and \ref{flux}~(b). Here, we
assume that the Doppler velocity of the IRIS jets can reach the
maximum speed of $\sim$150~km~s$^{-1}$ during their whole lifetime
($\sim$110~s), as shown by two vertical ticks in
Figure~\ref{flux}~(b). Because the IRIS observational mode is
scanned, we could not obtain the entire evolution of a fixed
position. Therefore, the kinetic energy in the transition region
estimated here is an upper-limit value, which is about
2$\times$10$^{26}$~erg.

\begin{figure}
\includegraphics[width=\columnwidth]{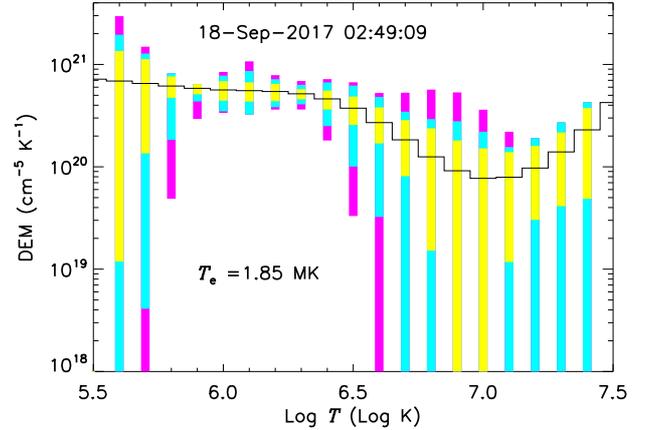}
\caption{DEM analysis result. The black profile shows the
best-fitted DEM curve from AIA observations. The yellow, turquoise,
and purple rectangles represent the regions that contain 50\%,
51\%$-$80\%, and 81\%$-$95\% of the Monte Carlo solutions.}
\label{dem}
\end{figure}

We then calculate the upper limit \citep{Aschwanden05} of thermal
energy (E$_t~\approx~3 n_e k_B T_e l^3$) in the corona from the EUV
brightening. Here, a cube volume ($l^3$) is assumed for the EUV
brightening, $n_e$ is the typical coronal number density
\citep[$\sim$10$^9$~cm$^{-3}$,][]{Tian16}, $k_B$ is Boltzmann
constant, $l$ is estimated from the scale ($\sim$3\arcsec) of the
EUV brightening in Figure~\ref{magnetic}. $T_e$ can be obtained from
the differential emission measure (DEM) result using SDO/AIA data in
six EUV channels \citep{Cheng12} at the peak brightness, which is
about 1.85~MK, as shown in Figure~\ref{dem}. Notice that the
confident temperature (log T) range is from 5.8 to 6.6, given on the
uncertainties (colored rectangles) of the DEM analysis. Thus the
upper limit of the thermal energy is around 7$\times$10$^{24}$~erg.
Our results suggest that the released magnetic energy in the
photosphere is big enough to drive the IRIS jets in the chromosphere
and transition region, and then cause the EUV brightening in the
corona, in the case that the thermal energy produced in the
transition region is of the same order (10$^{26}$~erg) of the E$_k$,
i.e., $E_m~\gg~E_k~+~E_t$.

\begin{figure}
\includegraphics[width=\columnwidth]{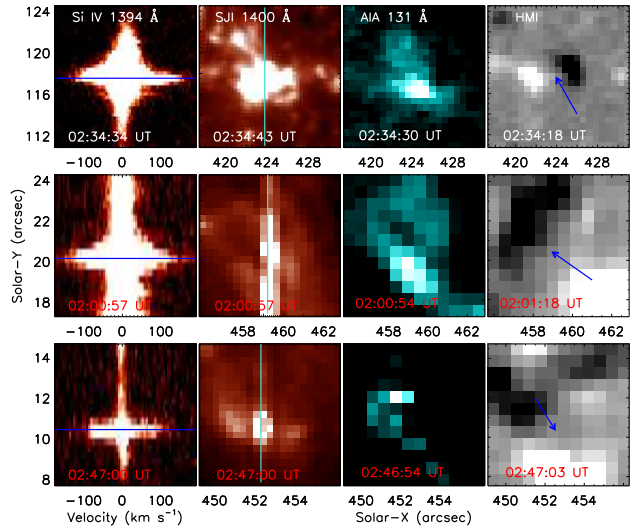}
\caption{Three similar reconnection events during this IRIS
observation.} \label{more}
\end{figure}

One reconnection event is detailed studied in the previous section,
it originates in the photosphere and subsequently spreads to the
corona. Then more cases from the IRIS and SDO observations are
found. Figure~\ref{more} shows three similar reconnection events in
this IRIS FOV during its observational time. These reconnection
events exhibit the same features. They are firstly identified as the
IRIS jets with fast bi-directional velocities in the transition
region (first column), such as the simultaneous enhancements of two
extended line wings of \siiv~1393.76~{\AA}. They all display the
enhanced coronal emission, i.e., AIA 131~{\AA} brightening (third
column). Finally, all these events are located above the polarity
inversion line between the positive and negative magnetic fields, as
indicated by the blue arrows. Although there are a lot of
transition-region and coronal enhanced events in the imaging
observations, most of them are missed by the slit of IRIS.
Therefore, the observations of IRIS jets with fast bi-directional
velocities are rare. Likely, four such IRIS jets are observed, and
the underlying photospheric magnetic field shows flux cancellation,
while the overlying coronal emission appears EUV brightening, which
confirm the process of magnetic reconnection from the photosphere
through the chromosphere and transition region to the corona. All
our observations suggest that the IRIS jets driven by the
photospheric magnetic fields could light up the corona. However, due
to the limitation of IRIS observations, at what percent this kind of
reconnection event contributes to the coronal heating is hard to be
estimated.

\begin{figure}
\includegraphics[width=\columnwidth]{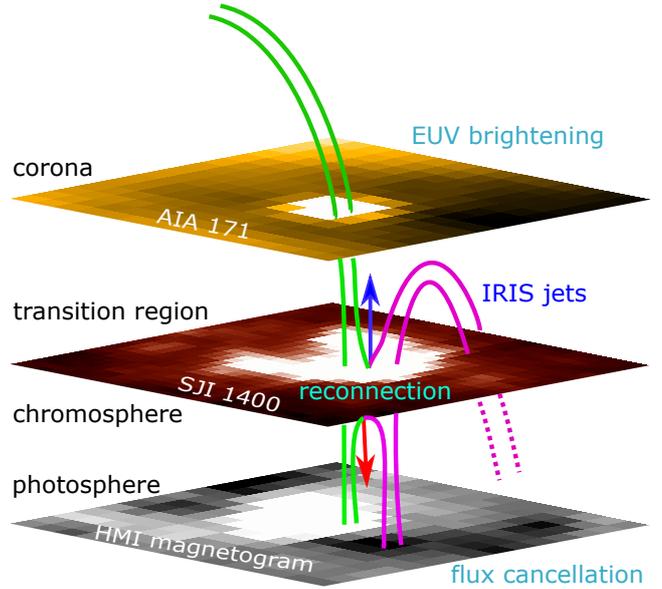}
\caption{Sketch plot of the reconnection events. The blue and red
arrows indicate the IRIS jets with bi-directions along LOS in the
chromosphere and transition region. The purple and green curves
represent the magnetic field lines.} \label{sketch}
\end{figure}

To illustrate the process of magnetic reconnection, a sketch is
showed in Figure~\ref{sketch}. They are firstly identified as the
IRIS jets with fast bi-directional velocities in NUV and FUV lines
(blue and red arrows in the middle panel), indicating that the
reconnection sites are ranging from the lower chromosphere (\mgii~k)
to the transition region (\siiv). On the other hand, the
reconnection events originate in the photosphere, which exhibit the
flux cancellation between positive and negative magnetic fields, as
shown in the lower panel. They can spread to the corona and cause
the EUV brightening, as shown in the upper panel. Our observations
of reconnection events agree well with the 2-D reconnection model
\citep{Shibata99,Priest00}, the magnetic reconnection occurs at the
X-point sites (i.e., chromosphere and transition region) where
anti-parallel magnetic field lines converge and reconnect, and these
magnetic field lines are rooted in the positive and negative fields,
as shown by the purple and green curves in Figure~\ref{sketch}. The
IRIS jets could be the reconnection outflows along the LOS, which
are driven by the flux cancellation in the photosphere and can cause
the EUV brightening in the corona.

There have been lots of observations of transient brightenings in
the chromosphere, transition region or corona in the past four
decades, i.e., the High-Resolution Telescope and Spectrograph (HRTS)
rocket flights from 1979 \citep{Dere81}, such as coronal bullets,
HRTS jets, transition-region explosive events, blinkers, nanoflares,
spicules, and solar tornadoes. Sometimes, they might be associated
with changing magnetic field conditions in the photosphere. However,
these small-scale events were only detected in one individual or two
solar atmospheres at each observation, and their observations were
separated to spectroscopy or image. For example, coronal bullets
were observed as small and discrete ejecta of chromospheric plasmas
in EUV lines, their speed could accelerate to be
$\sim$450~km~s$^{-1}$ \citep{Brueckner80,Karpen84}. HRTS jets
changed their blueshifted and redshifted components rapidly, and the
derived LOS velocities were in the range of 10$-$100~km~s$^{-1}$
\citep{Karpen82,Brueckner83}. Transition-region explosive events
were identified as non-Gaussian broadening profiles in EUV lines
\citep{Dere91,Innes13}. Blinkers were detected as the intensity
enhancements in EUV/SXR lines \citep{Harrison97,Chae00}. Nanoflares
were observed as EUV brightenings in the corona
\citep{Benz98,Ishikawa17}. Spicules were dynamic jet-like features
in the chromosphere or corona \citep{Pereira16,Dep17}. Solar
tornadoes were observed as tornado-like structures of magnetized
plasmas in the corona \citep{Su12,Wedemeyer12}. Using the
spectroscopic (IRIS) and imaging (SDO/AIA and SDO/HMI) observations,
the reconnection events in this study can be observed in almost all
the solar atmospheres, i.e., flux cancellation in the photosphere,
IRIS jets in the chromosphere and transition region, and EUV
brightening in the corona, as shown in the Figure~\ref{sketch}.

\section{Summary}
Using the spectroscopic and imaging observations from IRIS and SDO,
we detailed report a reconnection event that firstly identify as
IRIS jets in the chromosphere and transition region, which are
associated with the magnetic flux cancellation in the photosphere
and the EUV brightening in the corona. The IRIS jets can be detected
in the spectral lines of \mgii~k, \cii\ and \siiv, implying that the
site of magnetic reconnection is from the lower chromosphere to the
transition region. The contemporaneous flux cancellation confirms
the magnetic reconnection, which also indicates that the driver of
the energy release lies in the photosphere. While the AIA EUV
brightening suggests that the IRIS jets driven by the underlying
flux cancellation can cause the overlying coronal emission. Then
another three reconnection events are found during this IRIS
observation, which further suggest that the origin of coronal
heating lies in the photosphere and support the heating model of
small-scale magnetic reconnection. Future works will focus on
analyzing much more cases and performing a statistical research.

\section*{Acknowledgements}
The authors would like to thank the anonymous referee for his
valuable comments. IRIS is a NASA small explorer mission developed
and operated by LMSAL with mission operations executed at NASA Ames
Research center and major contributions to downlink communications
funded by ESA and the Norwegian Space Centre. SDO is a mission of
NASA's Living With a Star Program, a program designed to understand
the causes of solar variability and its impacts on Earth. We thank
Drs. Tie~Liu and Lei~Lu for the sketch plot. This work is supported
by NSFC under grants 11603077, 11573072, 11673034, 11533008,
11773079, 11790302, 11333009, the CRP (KLSA201708), the Youth Fund
of Jiangsu Nos. BK20161095, and BK20171108, as well as National
Natural Science Foundation of China (U1731241), the Strategic
Priority Research Program on Space Science, CAS, Grant No.
XDA15052200 and XDA15320301. D.~Li is supported by the Specialized
Research Fund for State Key Laboratories. The Laboratory No.
2010DP173032.


\bsp
\end{document}